\documentclass[11pt]{amsart}
\pdfoutput=1

\usepackage{latexsym,natbib, color}
\usepackage{graphicx}

\usepackage{amsmath, fullpage, booktabs}
\usepackage[colorlinks=TRUE, citecolor=blue]{hyperref}

\allowdisplaybreaks

\newtheorem{Theorem}{Theorem}

\newcommand{\Q}{{\mathbf{Q}}}
\newcommand{\PP}{{\mathbf{P}}}
\newcommand{\pii}{{\boldsymbol\pi}}
\newcommand{\X}{{\mathbf{X}}}
\newcommand{\St}{{\mathbf{S}}}
\newcommand{\dif}{\mathrm{d}}

\begin{document}

\title{On Estimation for Brownian Motion Governed by Telegraph Process with Multiple Off States}

\author[V. Pozdnyakov, L.M. Elbroch, C. Hu,  T. Meyer,  and J. Yan]{Vladimir Pozdnyakov$^1$$^*$, L. Mark Elbroch$^2$, Chaoran Hu$^1$,  Thomas Meyer$^4$,  and Jun Yan$^{1,3}$}

\thanks{\\
1. Department of Statistics, University of Connecticut, 215 Glenbrook Road, Storrs, CT 06269-4120\\
2. Panthera, 8 West 40th Street, 18th Floor, NY, NY 10018\\
3. Center for Environmental Sciences and Engineering, University of Connecticut, 3107 Horsebarn Hill Road, Storrs, Connecticut 06269-4210\\
4. Department of Natural Resources and the Environment, University of Connecticut, 1376 Storrs Road, Storrs, Connecticut 06269-4087\\
* E-mail: {\tt vladimir.pozdnyakov@uconn.edu}}

\begin{abstract}
Brownian motion whose infinitesimal variance changes according to a
three-state continuous time Markov Chain is studied. This Markov Chain
can be viewed as a telegraph process with one on state and two off
states. We first derive the distribution of occupation time of the on
state. Then  the result is used to develop a likelihood estimation
procedure when the stochastic process at hand is observed at discrete,
possibly irregularly spaced time points. The likelihood function is
evaluated with the forward algorithm in the general framework of
hidden Markov models. The analytic results are confirmed with
simulation studies. {  The estimation procedure is applied
  to analyze the position data from a mountain lion.}



\end{abstract}

\maketitle

\section{Introduction}

{
Random walks on a plane, whether simple, biased, or correlated, have a
long history of being employed by ecologists to model the movement of
animals, micro-organisms, and cells on a small time scale.
By the functional Central Limit Theorem, from an appropriate distance
any random walk (under some mild regularity conditions) looks like a
Brownian Motion (BM). So, it is not surprising that recently
diffusions are often used to model animal movement on a large time
scale \citep[e.g.,][]{Prei:etal:mode:2004, Tilles:Petr:2016}.
An excellent review on applications of random walks and diffusions in
this area of research can be found in \citet{Codling:etal:2008}.

\citet{Horn:etal:2007} introduced the Brownian bridge movement model
(BBMM) that, in essence, assumes that animal movement is perpetual and
described by a BM\@. Pauses in animal movement (on a small time scale)
were first introduced in \citet{Othmer:1988} where the dispersal of
cells or organisms is modeled by a process that comprises a sequence
of alternating pauses and jumps.
The moving-resting process introduced in \citet{Yan:etal:2014} and further
investigated in \citet{Pozd:etal:2017} allows an animal to have two states,
moving and resting. In the moving state, the motion is characterized by a BM;
in the resting state, there is no movement. The duration in either
moving or resting states is assumed to be exponentially distributed.

Properties and fitting of the moving-resting model are based on
results for telegraph processes (the alternating renewal process or
the on-off process) that were obtained in
\citet{Perr:Stad:Zack:firs:1999}, \citet{DiCrescenzo:2001},
\citet{Stad:Zack:tele:2004}, and \citet{Zack:gene:2004}.
The distribution of total time spent in a state plays a critical role in
applications driven by a telegraph process \citep{Zacks:2012}.
In particular, a BM governed by a telegraph process is an active area of research such as
being recently employed in continuous-time option
pricing theory \citep[e.g.,][]{DiCrescenzo:Pellerey:2002,
  Kolesnik:Ratanov:2013, DiCrescenzo:etal:2014,
  DiCrescenzo:Zhacs:2015}.

In animal movement ecology, it is reasonable to assume that there are
very different explanations for why a predator is not moving.
For example, an animal might spend time resting (as in \citet{Yan:etal:2014}),
consuming a prey item, or denning. Resting can be assumed to
not last even a single day. However, some predators that can kill a
(relatively) large prey item evolved highly elastic guts, and they consume the
kill by repeatedly gorging and digesting over a prolonged period
called {\it handling}.  For example, mountain lions
({\it Puma concolor}) might remain at a kill for days. Both resting and
handling are periodic in the time scales of this model but denning is
not, and it is inapplicable to male mountain lions in any case. Therefore,
this model concerns only two non-moving activities, resting and handling, and
it is clear that their durations must be different.

This observation motivates our model. In the new model we have one
moving state and two motionless states. From a motionless state one
always switches to the moving state. Nonetheless, when moving ends,
the motionless state type is chosen randomly. For tractability,
all the durations (or holding times) are exponentially distributed.
We will call this continuous-time process a {\it moving-resting-handling
  process}, or {\it MRH process}.
An extension of the telegraph process to an alternating process with
three states is studied in \citet{Bshouty:etal:2012}.
The difference is that in \citet{Bshouty:etal:2012} three states
alternate deterministically within a renewal cycle.
In our case we have only two states within a renewal cycle but one of
the motionless states is chosen at random.
}

In practice, a MRH process is typically observed at discrete, possibly
irregularly spaced time points. Estimation of MRH process parameters is
challenging because the states are unobserved, and the observed sequence is
not Markov. Our estimation procedure uses techniques developed for the hidden
Markov model (HMM). More specifically, the dynamic programming, or the
forward algorithm, for HMM is employed to construct the true likelihood
\citep[e.g.][]{Capp:etal:Infe:2005}. As will be seen, the key to this problem
is the distribution of the time that the MRH process spends in the moving
state. Our methodology differs from the standard approach to occupation time
distribution in continuous-time Markov chain \citep{Seri:2000}.
The method is general so that it remains valid when the holding times are not
exponentially distributed, in which case, the state process is semi-Markov;
see discussion in Section~\ref{sec:conc}.
{
An implementation of the methods in this paper is publicly available
in R package \texttt{smam} \citep{Rpkg:smam}.
}

\section{Formal Description of MRH Process}

Let $S(t)$, $t \geq 0$, be a continuous-time Markov Chain with the state space
$\{0,1,2\}$ and the transition rate matrix
\begin{equation}\label{transition*rate*matrix}
\Q=\begin{pmatrix}
    -\lambda_0 & \lambda_0 \, p_1 & \lambda _0 \, p_2 \\
     \lambda_1 & -\lambda_1 & 0 \\
     \lambda_2 & 0 & -\lambda_2 \\
  \end{pmatrix}
\end{equation}
where $p_1, p_2,\lambda_0,\lambda_1,\lambda_2>0$ and $p_1+p_2=1$.
The zero entries in the matrix means that state~1 or state~2 do not transit
between themselves; only a transition to state~0 is allowed from either of them.
In animal movement modeling, the mean duration in state~0, 1,~and~2 are,
respectively, $1/\lambda_0$, $1/\lambda_1$, and $1/\lambda_2$.
We assume that the initial distribution $\nu_0$ of $S(0)$ is stationary, that is,
\begin{equation}\label{stationary*distribution}
\nu_0=\pii=(\pi_0,\pi_1,\pi_2)=
\frac{1}{1/\lambda_0+p_1/\lambda_1+p_2/\lambda_2}
\left(\frac{1}{\lambda_0}, \frac{p_1}{\lambda_1}, \frac{p_2}{\lambda_2}\right).
\end{equation}
Recall that $\pii$ has to satisfy $0=\pii\Q$.

Let $B(t)$ be the standard BM independent of $S(t)$.
Then the MRH process is given by
\begin{equation}\label{MRH*process}
X(t)=\sigma \int_0^t1_{\{S(s)=0\}}\dif B(s),
\end{equation}
where $\sigma>0$ is an infinitesimal standard deviation.

Estimation of the MRH process parameters
${\boldsymbol\theta}=(\lambda_0,\lambda_1,\lambda_2,p_1,\sigma)$
is based on observations at discrete, possibly irregularly spaced time points.
The observed data are represented by the vector of observed changes in location
\begin{equation*}
\X = \big(X(t_1)-X(0), X(t_2)-X(t_1), \dots, X(t_n)-X(t_{n-1})\big),
\end{equation*}
where $0<t_1<\dots<t_n$ are the time points of the observations.
As mentioned earlier, the difficulty is that the MRH process itself is not
Markov. However, the location-state process $\{X(t), S(t)\}$ is Markov.
So, our first objective is to derive formulas for transitional probabilities
of the location-state process. The key random variable here is the total time
spent in state 0 in the time interval $[0,t]$:
\begin{equation}\label{time*spent*in*moving}
M(t)=\int_0^t1_{\{S(s)=0\}}\dif s.
\end{equation}
We also can call this random variable {\it 0-state occupation time} by
time~$t$.

A continuous-time Markov Chain can be alternatively described by representing
the process $S(t)$ as a combination of a discrete time Markov Chain, holding
times, and initial distribution~$\nu$. More specifically, let $p_{ij}$ be the
probability of switching to state~$j$ at the next jump given that
we are currently in state~$i$. The matrix
\begin{equation*}
\PP=(p_{ij})=\begin{pmatrix}
              0 & p_1 & p_2 \\
              1 & 0 & 0 \\
              1 & 0 & 0 \\
            \end{pmatrix}
\end{equation*}
is a stochastic matrix, and it is the transition matrix of the embedded
(discrete time) Markov Chain of process $S(t)$.
The time spent in a particular state $i$ between two consecutive jumps is
called the holding time. The holding time has exponential
distribution with rate $\lambda_i$. For our task this representation (via an
embedded Markov Chain and holding times) is a bit more convenient. Note also that in the case of the
standard telegraph process the associated stochastic matrix of the embedded Markov chain is
 \begin{equation*}
\begin{pmatrix}
0 & 1\\
1 & 0 \\
\end{pmatrix}.
\end{equation*}

Our technique is different from the general approach to the distribution of
occupation times in homogeneous finite-state Markov processes (e.g., \citet{Seri:2000}).
To develop computationally efficient estimation procedure we exploit the
specific structure of our Markov chain. More specifically, a telegraph process
can be associated with $S(t)$ if we collapse states~1 and~2 into one state.
For this new state the holding time is distributed as a mixture of two
exponential distributions. As a consequence, the telegraph process is not
Markov. This makes computing the likelihood function for $\X$ challenging,
because algorithms like the forward algorithm are not applicable. That is,
results for telegraph processes can not be directly employed, because we do
need to distinguish states 1 and 2. We use a certain periodicity of the Markov
Chain and extend the technique developed in \citet{DiCrescenzo:2001}
for telegraph processes to obtain the joint distribution of $M(t)$ and $S(t)$.
{  An alternative approach can be developed by extending the method
presented in \citet{Zacks:2012}.}

\section{Distribution of Occupation Time $M(t)$ Given $S(0)=0$}\label{M(t)*when*S(0)=0*section}

To simulate process $S(t)$ that starts with $S(0)=0$, we need the following
independent sequences of random variables:
\begin{enumerate}
\item $\{M_k\}_{k\geq 1}$ are independent identically distributed (iid) random variables with ${\rm Exp}(\lambda_0)$ distribution,
\item $\{R_k\}_{k\geq 1}$ are iid random variables with ${\rm Exp}(\lambda_1)$,
\item $\{H_k\}_{k\geq 1}$ are iid random variables with ${\rm Exp}(\lambda_2)$,
\item $\{\xi_k\}_{k\geq 1}$ are iid random variables with $P(\xi_k=1)=p_1$ and $P(\xi_k=0)=p_2$.
\end{enumerate}
Having these sequences defined we can proceed as follows.
To generate a particular realization of $S(t)$, first, generate $M_1$,
the time duration the process spends in state~0. Then generate $\xi_1$ to
decide whether it jumps to state~1 or~2. Depending on $\xi_1$ generate the
duration $R_1$ or $H_1$. After that, switch back to state~0, and so on.

Let us introduce some auxiliary random variables.
Let $U_k=\xi_kR_k+(1-\xi_k)H_k$, $C_k=M_k+U_k$, and
\begin{equation*}
N(t)=\sup\{n\geq 0:\sum_{k=1}^nC_k\leq t\}.
\end{equation*}
Here and everywhere in the text, by convention, a summation over an empty set
is 0, for instance, $\sum_{k=1}^0C_k=0$.
Random variable $N(t)$ is the number of full cycles $C_k$ by time $t$.

First, we consider the distribution of occupation time $M(t)$ when $S(t)=0$.
Denote $P_i(\cdot)=P(\cdot|S(0)=i)$, where $i=0,1,2$. With probability~1 the
random variable $M(t)\in [0,t]$, and it has an atom at $t$ in the following
sense:
\begin{equation*}
P_0(M(t)=t,S(t)=0)=P_0(M(t)=t)=P(M_1>t)=e^{-\lambda_0t}.
\end{equation*}
Now, fix $0<s<t$. Then we have
\begin{align*}
P_0(M(t)\in \dif s, S(t)=0)&=\sum_{n=0}^\infty P_0(M(t)\in \dif s,S(t)=0,N(t)=n)\\
                      &=\sum_{n=1}^\infty P_0(M(t)\in \dif s,S(t)=0,N(t)=n),
\end{align*}
because $S(0)=0$, $S(t)=0$ and $N(t)=0$ implies $M(t)=t$.

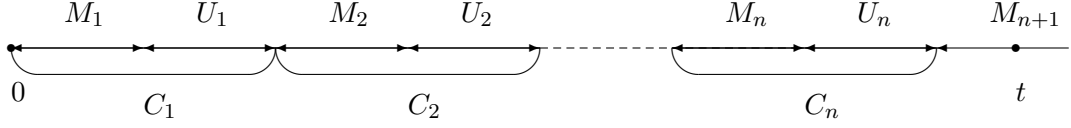
\begin{figure}[tbp]
\begin{center}
\begin{picture}(420, 100)
\put(10,50){\vector(1,0){50}}
\put(10,50){\circle*{3}}
\put(10,30){$0$}
\put(60,50){\vector(-1,0){50}}
\put(30,60){$M_1$}
\put(60,50){\vector(1,0){50}}
\put(110,50){\vector(-1,0){50}}
\put(80,60){$U_1$}
\put(60,50){\oval(100,20)[b]}
\put(60,25){$C_1$}
\put(110,50){\vector(1,0){50}}
\put(160,50){\vector(-1,0){50}}
\put(130,60){$M_2$}
\put(160,50){\vector(1,0){50}}
\put(210,50){\vector(-1,0){50}}
\put(180,60){$U_2$}
\put(160,50){\oval(100,20)[b]}
\put(160,25){$C_2$}
\multiput(210,50)(5,0){20}
{\line(1,0){3}}
\put(260,50){\vector(1,0){50}}
\put(310,50){\vector(-1,0){50}}
\put(280,60){$M_n$}
\put(310,50){\vector(1,0){50}}
\put(360,50){\vector(-1,0){50}}
\put(330,60){$U_n$}
\put(310,50){\oval(100,20)[b]}
\put(310,25){$C_n$}
\put(390,50){\circle*{3}}
\put(390,30){$t$}
\put(410,50){\vector(-1,0){50}}
\put(380,60){$M_{n+1}$}
\end{picture}
\end{center}
\caption{$S(t)=0$ and $N(t)=n$ given that $S(0)=0$.}
\label{fig:pic1}
\end{figure}

Next, for $n\geq 1$ we get, from Figure~\ref{fig:pic1}, that
\begin{align*}
&\quad P_0(M(t)\in \dif s,S(t)=0,N(t)=n)\\
&=P\left(\sum_{k=1}^nM_k+\sum_{k=1}^nU_k\leq t, \sum_{k=1}^{n+1}M_k+\sum_{k=1}^nU_k> t,t-\sum_{k=1}^nU_k\in \dif s\right)\\
&=P\left(\sum_{k=1}^nM_k\leq s, \sum_{k=1}^{n+1}M_k> s,\sum_{k=1}^nU_k\in t- \dif s\right)\\
&=P\left(\sum_{k=1}^nM_k\leq s, \sum_{k=1}^{n+1}M_k> s\right)P\left(\sum_{k=1}^nU_k\in t- \dif s\right)\\
&=\left[P\left(\sum_{k=1}^nM_k\leq s\right)- P\left(\sum_{k=1}^{n+1}M_k\leq s\right)\right]P\left(\sum_{k=1}^nU_k\in t- \dif s\right).
\end{align*}
Here we use independence of $\{M_k\}_{k\geq 1}$ and $\{U_k\}_{k\geq 1}$.

The sums $\sum_{k=1}^nM_k$ and $\sum_{k=1}^{n+1}M_k$ have gamma distributions,
${\rm Gamma}(n,\lambda_0)$ and ${\rm Gamma}(n+1,\lambda_0)$, respectively.
The distribution of $\sum_{k=1}^nU_k$ can be expressed in terms of the convolution
of gamma distributions. More specifically, by conditioning on
$\{\xi_k\}_{1\leq k\leq n}$ one can show that
\begin{equation*}
P\left(\sum_{k=1}^nU_k\leq s\right)=\sum_{k=0}^nP\left(\sum_{j=1}^kR_j+\sum_{j=1}^{n-k}H_j\leq s\right) {n \choose k} p_1^kp_2^{n-k}.
\end{equation*}
Random variables $\sum_{j=1}^kR_j$ and  $\sum_{j=1}^{n-k}H_j$ are independent,
and they have ${\rm Gamma}(k,\lambda_1)$ and ${\rm Gamma}(n-k,\lambda_2)$
distributions, respectively. For the convolution of gamma distributions, we
refer the reader to \citet{Math:1982} and \citet{Mosc:1985}.

Next, let us work out the case when $S(t)=1$. Again, the random variable
$M(t)\in [0,t]$, but now it has no atoms. For any  $0<s<t$, we have
\begin{align*}
P_0(M(t)\in \dif s, S(t)=1)&=\sum_{n=0}^\infty P_0(M(t)\in \dif s,S(t)=1,N(t)=n)
\end{align*}

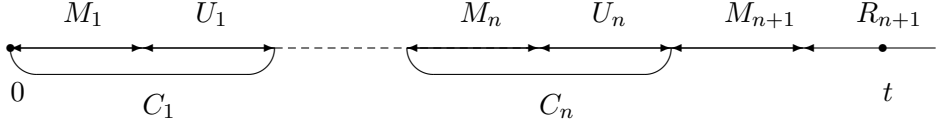
\begin{figure}[tbp]
\begin{center}
\begin{picture}(370, 100)
\put(10,50){\vector(1,0){50}}
\put(10,50){\circle*{3}}
\put(10,30){$0$}
\put(60,50){\vector(-1,0){50}}
\put(30,60){$M_1$}
\put(60,50){\vector(1,0){50}}
\put(110,50){\vector(-1,0){50}}
\put(80,60){$U_1$}
\put(60,50){\oval(100,20)[b]}
\put(60,25){$C_1$}
\multiput(110,50)(5,0){20}
{\line(1,0){3}}
\put(160,50){\vector(1,0){50}}
\put(210,50){\vector(-1,0){50}}
\put(180,60){$M_n$}
\put(210,50){\vector(1,0){50}}
\put(260,50){\vector(-1,0){50}}
\put(230,60){$U_n$}
\put(210,50){\oval(100,20)[b]}
\put(210,25){$C_n$}
\put(260,50){\vector(1,0){50}}
\put(310,50){\vector(-1,0){50}}
\put(280,60){$M_{n+1}$}
\put(340,50){\circle*{3}}
\put(340,30){$t$}
\put(360,50){\vector(-1,0){50}}
\put(330,60){$R_{n+1}$}
\end{picture}
\end{center}
\caption{$S(t)=1$ and $N(t)=n$ given that $S(0)=0$.}
\label{fig:pic2}
\end{figure}

Then for $n\geq 0$ we get, from Figure~\ref{fig:pic2}, that
\begin{align*}
&\quad P_0(M(t)\in \dif s, S(t)=1,N(t)=n)\\
&=P\left(\sum_{k=1}^{n+1}M_k+\sum_{k=1}^nU_k\leq t, \sum_{k=1}^{n+1}M_k+\sum_{k=1}^nU_k+R_{n+1}> t,\sum_{k=1}^{n+1}M_k\in \dif s, \xi_{n+1}=1\right)\\
&=P\left(\sum_{k=1}^nU_k\leq t-s, \sum_{k=1}^{n}U_k+R_{n+1}> t-s,\sum_{k=1}^{n+1}M_k\in \dif s,\xi_{n+1}=1\right)\\
&=p_1P\left(\sum_{k=1}^nU_k\leq t-s, \sum_{k=1}^{n}U_k+R_{n+1}> t-s\right)P\left(\sum_{k=1}^{n+1}M_k\in \dif s\right)\\
&=p_1\left[P\left(\sum_{k=1}^nU_k\leq t-s\right)- P\left(\sum_{k=1}^{n}U_k+R_{n+1}\leq t-s\right)\right]P\left(\sum_{k=1}^{n+1}M_k\in \dif s\right).
\end{align*}
Random variable $\sum_{k=1}^{n+1}M_k$ has ${\rm Gamma}(n+1,\lambda_0)$ distribution. As before,
\begin{equation*}
P\left(\sum_{k=1}^nU_k\leq s\right)=\sum_{k=0}^nP\left(\sum_{j=1}^kR_j+\sum_{j=1}^{n-k}H_j\leq s\right) {n \choose k} p_1^kp_2^{n-k},
\end{equation*}
and
\begin{equation*}
P\left(\sum_{k=1}^nU_k+R_{n+1}\leq s\right)=\sum_{k=0}^nP\left(\sum_{j=1}^{k+1}R_j+\sum_{j=1}^{n-k}H_j\leq s\right) {n \choose k} p_1^kp_2^{n-k}.
\end{equation*}

To summarize our findings let us first introduce the following notation:
\begin{enumerate}
\item $G(x,\alpha,\beta)$, where $\alpha\geq 0,\beta>0$, is the cdf of ${\rm
    Gamma}(\alpha,\beta)$ distribution; by convention, ${\rm Gamma}(0,\beta)$
  distribution is the degenerate distribution with atom 1 at 0;
\item $g(x,\alpha,\beta)$, where $\alpha,\beta>0$, is the pdf of ${\rm
    Gamma}(\alpha,\beta)$ distribution;
\item $F(x,\alpha_1,\beta_1,\alpha_2,\beta_2)$, where $\alpha_1,\alpha_2\geq
  0,\beta_1,\beta_2>0$, is the cdf of the convolution of ${\rm
    Gamma}(\alpha_1,\beta_1)$ and ${\rm Gamma}(\alpha_2,\beta_2)$;
  note that, for example, $F(x,0,\beta_1,\alpha_2,\beta_2)\equiv
  G(x,\alpha_2,\beta_2)$;
\item $f(x,\alpha_1,\beta_1,\alpha_2,\beta_2)$, where $\beta_1,\beta_2>0$,
  $\alpha_1,\alpha_2\geq 0$, and $\alpha_1+\alpha_2>0$, is the pdf of
  $F(x,\alpha_1,\beta_1,\alpha_2,\beta_2)$;
\item $H(x, \alpha_1, \beta_1, \alpha_2, \beta_2) =
  F(x, \alpha_1, \beta_1, \alpha_2, \beta_2)
  - F(x, \alpha_1 + 1, \beta_1, \alpha_2, \beta_2)$, where $\beta_1,\beta_2>0$,
  $\alpha_1,\alpha_2\geq 0$, and $\alpha_1+\alpha_2>0$, is the difference in
  cdf with parameters only differing by $\alpha_1$ versus $\alpha_1 + 1$.
\end{enumerate}

Finally, let us denote the (defective) densities of $M(t)$ as
\begin{equation}\label{M(t)*density}
p_{ij}(s,t)=P_i(M(t)\in \dif s,S(t)=j)/\dif s,
\end{equation}
where $t\geq 0$, $0<s<t$, $i,j=0,1,2$.

Here is the main result of the section.
\begin{Theorem}\label{thm:S0=0}
Let $t\geq 0$ and $0<s<t$. Then
\begin{equation}\label{P0}
P_0(M(t)=t,S(t)=0)=e^{-\lambda_0t},
\end{equation}
and the densities are given by
\begin{equation}\label{p00}
p_{00}(s,t)=\sum_{n=1}^\infty\left[ G(s,n,\lambda_0)-G(s,n+1,\lambda_0)\right]\sum_{k=0}^nf(t-s,k,\lambda_1,n-k,\lambda_2) {n\choose k}p_1^kp_2^{n-k},
\end{equation}
\begin{equation}\label{p01}
p_{01}(s,t)=\sum_{n=0}^\infty p_1g(s,n+1,\lambda_0)\sum_{k=0}^n H(t-s,k,\lambda_1,n-k,\lambda_2) {n\choose k}p_1^kp_2^{n-k},
\end{equation}
and
\begin{equation}\label{p02}
p_{02}(s,t)=\sum_{n=0}^\infty p_2g(s,n+1,\lambda_0)\sum_{k=0}^n H(t-s,k,\lambda_2,n-k,\lambda_1) {n\choose k}p_2^kp_1^{n-k}.
\end{equation}
\end{Theorem}

Note that the last formula of Theorem~\ref{thm:S0=0} can be obtained from the
previous one by interchanging state~1 and state~2.

\section{Distribution of Occupation Time $M(t)$ Given $S(0)=1$}\label{M(t)*when*S(0)=1*section}

Let $\{M_k\}_{k\geq 1}$,  $\{R_k\}_{k\geq 1}$, $\{H_k\}_{k\geq 1}$,
$\{\xi_k\}_{k\geq 1}$, and  $\{U_k\}_{k\geq 1}$ be the same sequences of
random variables as in Section~\ref{M(t)*when*S(0)=0*section}.
Let $R_0$ be an independent-of-everything random variable with
${\rm Exp}(\lambda_1)$ distribution. When $S(0)=1$, the sequence of holding
times starts from $R_0$; that is, we have: $R_0,M_1,U_1,M_2,U_2,\dots$.
This requires us to modify the definition of cycles. Now $C_1=R_0+M_1$, and
$C_k=U_{k-1}+M_k$ for $k\geq 1$. As before, the random variable $N(t)$ is the
number of cycles in time interval $[0,t]$:
\begin{equation*}
N(t)=\sup\{n\geq 0:\sum_{k=1}^nC_k\leq t\}.
\end{equation*}

Let us first consider the distribution of $M(t)$ when $S(t)=1$. Again, in this
case there is an atom, but now the atom is at $s=0$:
\begin{equation*}
P_1(M(t)=0,S(t)=1)=P_1(M(t)=0)=P(R_0>t)=e^{-\lambda_1t}.
\end{equation*}
Fix $0<s<t$. First note that $S(0)=1$, $S(t)=1$, and $N(t)=0$ implies that $M(t)=0$, therefore,
\begin{align*}
P_1(M(t)\in \dif s, S(t)=1)&=\sum_{n=0}^\infty P_1(M(t)\in \dif s,S(t)=1,N(t)=n)\\
                       &=\sum_{n=1}^\infty P_1(M(t)\in \dif s,S(t)=1,N(t)=n).
\end{align*}

\begin{figure}[tbp]
\centering
\begin{picture}(420, 100)
\put(10,50){\vector(1,0){50}}
\put(10,50){\circle*{3}}
\put(10,30){$0$}
\put(60,50){\vector(-1,0){50}}
\put(30,60){$R_0$}
\put(60,50){\vector(1,0){50}}
\put(110,50){\vector(-1,0){50}}
\put(80,60){$M_1$}
\put(60,50){\oval(100,20)[b]}
\put(60,25){$C_1$}
\put(110,50){\vector(1,0){50}}
\put(160,50){\vector(-1,0){50}}
\put(130,60){$U_1$}
\put(160,50){\vector(1,0){50}}
\put(210,50){\vector(-1,0){50}}
\put(180,60){$M_2$}
\put(160,50){\oval(100,20)[b]}
\put(160,25){$C_2$}
\multiput(210,50)(5,0){20}
{\line(1,0){3}}
\put(260,50){\vector(1,0){50}}
\put(310,50){\vector(-1,0){50}}
\put(280,60){$U_{n-1}$}
\put(310,50){\vector(1,0){50}}
\put(360,50){\vector(-1,0){50}}
\put(330,60){$M_n$}
\put(310,50){\oval(100,20)[b]}
\put(310,25){$C_n$}
\put(390,50){\circle*{3}}
\put(390,30){$t$}
\put(410,50){\vector(-1,0){50}}
\put(380,60){$R_{n}$}
\end{picture}
\caption{$S(t)=1$ and $N(t)=n$ given that $S(0)=1$.}
\label{fig:pic3}
\end{figure}
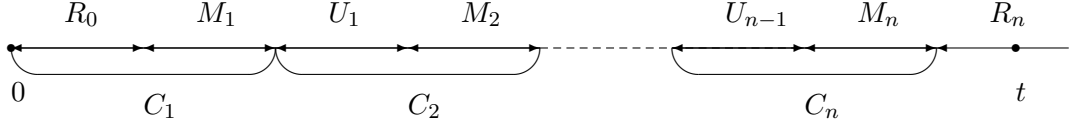

Finally, for $n\geq 1$ one can show (see Figure~\ref{fig:pic3}) that
\begin{align*}
&\quad P_1(M(t)\in \dif s,S(t)=1,N(t)=n)\\
&=P\left(R_0+\sum_{k=1}^{n}M_k+\sum_{k=1}^{n-1}U_k\leq t, R_0+\sum_{k=1}^{n}M_k+\sum_{k=1}^{n-1}U_k+R_n> t,\sum_{k=1}^{n}M_k\in \dif s, \xi_{n}=1\right)\\
&=P\left(R_0+\sum_{k=1}^{n-1}U_k\leq t-s, R_0+\sum_{k=1}^{n-1}U_k+R_n> t-s,\sum_{k=1}^{n}M_k\in \dif s, \xi_{n}=1\right)\\
&=p_1P\left(R_0+\sum_{k=1}^{n-1}U_k\leq t-s, R_0+\sum_{k=1}^{n-1}U_k+R_n> t-s\right)P\left(\sum_{k=1}^{n}M_k\in \dif s\right)\\
&=p_1\left[P\left(R_0+\sum_{k=1}^{n-1}U_k\leq t-s\right)-P\left(R_0+\sum_{k=1}^{n-1}U_k+R_n> t-s\right)\right]P\left(\sum_{k=1}^{n}M_k\in \dif s\right)\\
&=p_1g(s,n,\lambda_0)\\
&\times\sum_{k=0}^{n-1}\left[F(t-s,k+1,\lambda_1,n-1-k,\lambda_2)-F(t-s,k+2,\lambda_1,n-1-k,\lambda_2)\right] {n-1\choose k}p_1^kp_2^{n-1-k}ds\\
&=p_1g(s,n,\lambda_0) \times\sum_{k=0}^{n-1} H(t-s,k+1,\lambda_1,n-1-k,\lambda_2) {n-1\choose k}p_1^kp_2^{n-1-k}ds.
\end{align*}

The next case is when $S(t)=2$. In this situation $M(t)$ does not have atoms,
because we cannot switch from state~1 to state~2 without visiting state~0.
Since event $\{S(0)=1$, $S(t)=2$, and $N(t)=0\}$ is impossible, for $0<s<t$ we
have
\begin{align*}
P_1(M(t)\in \dif s, S(t)=2)&=\sum_{n=0}^\infty P_1(M(t)\in \dif s,S(t)=2,N(t)=n)\\
                       &=\sum_{n=1}^\infty P_1(M(t)\in \dif s,S(t)=2,N(t)=n),
\end{align*}
and for $n\geq 1$
\begin{align*}
&\ P_1(M(t)\in \dif s,S(t)=2,N(t)=n)\\
=&\ P\left(R_0+\sum_{k=1}^{n}M_k+\sum_{k=1}^{n-1}U_k\leq t, R_0+\sum_{k=1}^{n}M_k+\sum_{k=1}^{n-1}U_k+H_n> t,\sum_{k=1}^{n}M_k\in \dif s, \xi_{n}=0\right)\\
=&\ P\left(R_0+\sum_{k=1}^{n-1}U_k\leq t-s, R_0+\sum_{k=1}^{n-1}U_k+H_n> t-s,\sum_{k=1}^{n}M_k\in \dif s, \xi_{n}=0\right)\\
=&\ p_2P\left(R_0+\sum_{k=1}^{n-1}U_k\leq t-s, R_0+\sum_{k=1}^{n-1}U_k+H_n> t-s\right)P\left(\sum_{k=1}^{n}M_k\in \dif s\right)\\
=&\ p_2\left[P\left(R_0+\sum_{k=1}^{n-1}U_k\leq t-s\right)-P\left(R_0+\sum_{k=1}^{n-1}U_k+H_n> t-s\right)\right]P\left(\sum_{k=1}^{n}M_k\in \dif s\right)\\
=&\ p_2g(s,n,\lambda_0)\\
&\times\sum_{k=0}^{n-1}\left[F(t-s,k+1,\lambda_1,n-1-k,\lambda_2)-F(t-s,k+1,\lambda_1,n-k,\lambda_2)\right] {n-1\choose k}p_1^kp_2^{n-1-k}ds\\
=&\ p_2g(s,n,\lambda_0) \times\sum_{k=0}^{n-1} H(t-s, n-1-k,\lambda_2, k+1,\lambda_1) {n-1\choose k}p_1^kp_2^{n-1-k}ds.
\end{align*}

Finally, let us consider the case $S(t)=0$. Again, there are no atoms. For $0<s<t$
\begin{align*}
P_1(M(t)\in \dif s, S(t)=0)&=\sum_{n=0}^\infty P_1(M(t)\in \dif s,S(t)=0,N(t)=n),
\end{align*}
and for  $n\geq 0$
\begin{align*}
P_1(M(t)&\in \dif s,S(t)=0,N(t)=n)\\
                           &=P\left(R_0+\sum_{k=1}^{n}M_k+\sum_{k=1}^{n}U_k\leq t, R_0+\sum_{k=1}^{n+1}M_k+\sum_{k=1}^{n}U_k> t,t-\sum_{k=1}^{n}U_k-R_0\in \dif s\right)\\
                           &=P\left(\sum_{k=1}^{n}M_k\leq s, \sum_{k=1}^{n+1}M_k>s,\sum_{k=1}^{n}U_k+R_0\in t- \dif s\right)\\
                           &=P\left(\sum_{k=1}^{n}M_k\leq s, \sum_{k=1}^{n+1}M_k>s\right)P\left(\sum_{k=1}^{n}U_k+R_0\in t- \dif s\right)\\
                           &=\left[P\left(\sum_{k=1}^{n}M_k\leq s\right)-P\left(\sum_{k=1}^{n+1}M_k\leq s\right)\right]P\left(\sum_{k=1}^{n}U_k+R_0\in t- \dif s\right)\\
                           &=\left[G(s,n,\lambda_0)-G(s,n+1,\lambda_0)\right]\sum_{k=0}^{n}\left[f(t-s,k+1,\lambda_1,n-k,\lambda_2)\right] {n\choose k}p_1^kp_2^{n-k}ds.
\end{align*}

Thus, we have the following result.

\begin{Theorem}\label{thm:S0=1}
Let $t\geq 0$ and $0<s<t$. Then
\begin{equation}\label{P1}
P_1(M(t)=0,S(t)=1)=e^{-\lambda_1t},
\end{equation}
and the densities are given by
\begin{equation}\label{p10}
p_{10}(s,t)=\sum_{n=0}^\infty\left[G(s,n,\lambda_0)-G(s,n+1,\lambda_0)\right]\sum_{k=0}^{n}\left[f(t-s,k+1,\lambda_1,n-k,\lambda_2)\right] {n\choose k}p_1^kp_2^{n-k},
\end{equation}
\begin{equation}\label{p11}
p_{11}(s,t)=\sum_{n=1}^\infty p_1g(s,n,\lambda_0) \times\sum_{k=0}^{n-1} H(t-s,k+1,\lambda_1,n-1-k,\lambda_2) {n-1\choose k}p_1^kp_2^{n-1-k},
\end{equation}
and
\begin{equation}\label{p12}
p_{12}(s,t)=\sum_{n=1}^\infty p_2g(s,n,\lambda_0) \times\sum_{k=0}^{n-1} H(t-s,n-1-k,\lambda_2,k+1,\lambda_1) {n-1\choose k}p_1^kp_2^{n-1-k}.
\end{equation}
\end{Theorem}

In order to get densities $p_{2j}(s,t)$, $j=0,1,2$ we simply need to
interchange state~1 and state~2 in all the formulas of Theorem~\ref{thm:S0=1}.
Also let us note that Theorems~\ref{thm:S0=0} and~\ref{thm:S0=1} can be easily
extended to the case when there are more than two motionless states.

\section{Numerical Verification}
{

\begin{figure}[tbp]
\centering
\includegraphics[angle=0]{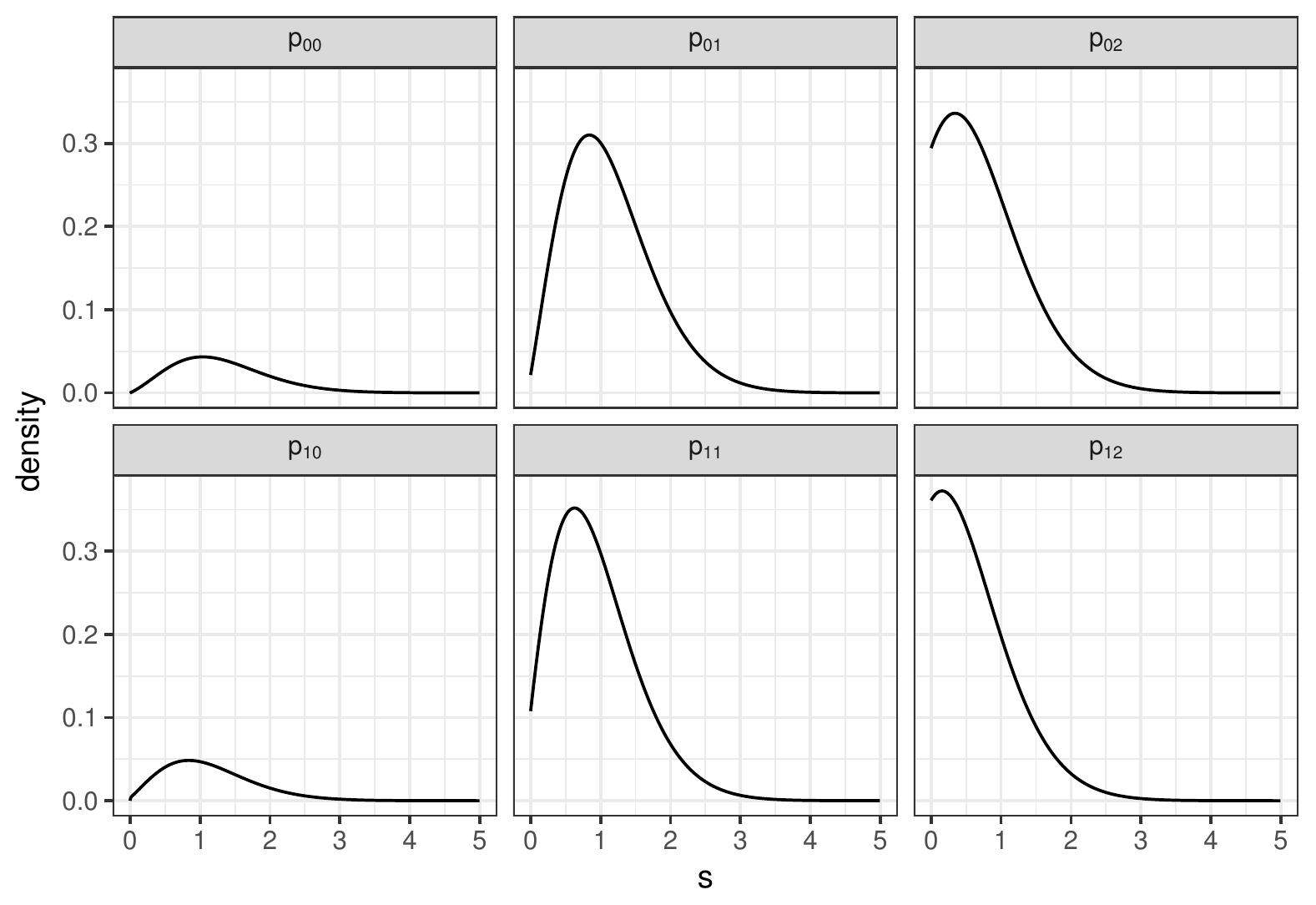}
\caption{Defective densities $p_{ij}(s,t)$: Theorem~\ref{thm:S0=0} and Theorem~\ref{thm:S0=1}}
\label{fig:mrh_density}
\end{figure}

Figure~\ref{fig:mrh_density} presents defective densities $p_{ij}(s,t)$
for two cases when the Markov Chain starts in state~0 and state~1.
The following model parameters are used:
$\lambda_0=4$, $\lambda_1=.5$, $\lambda_2=.1$, $p_1=.8$, and $t=10$.
Note that the total probability in both cases is slightly less than 1.
When the Markov Chain starts in state~0, the total probability adds up
to $1-e^{-40}$, because $M(10)$ has an atom at $s=10$.  When the Markov Chain
starts in state~1, the probability of the atom at $s=0$ is
relatively larger: $e^{-5}$. Because densities $p_{0j}(s,t)$ correspond
to the case when at $s=0$ the state process is in state 0, these occupation times
are longer on average than $p_{1j}(s,t)$.

Applications of the formulas in practice depends on how accurately the
infinite sums can be implemented. To check the accuracy of the implementation
and to verify that our formulas in Theorem~\ref{thm:S0=0} and
Theorem~\ref{thm:S0=1} are free of errors or typos, we simulated 1,000,000
realizations of the Markov chain $S(\cdot)$ for each theorem. The empirical
densities follow theoretical ones extremely closely (not shown).

We also performed another check.
There are two cases when the MHR process collapses to the moving-resting
process investigated in  \citet{Yan:etal:2014}. If $p_1$ is equal to 0 or 1,
then the MHR process (after the first visit of moving state) will alternate
only between two states. The other case is when $\lambda_1=\lambda_2$. The MHR
process will hit all three states, but state 1 and state 2 are
undistinguishable. This can be verified analytically. For example, one
can show that our formula~\eqref{p00} will simplify to the first term of (2.3)
in \citet{Zack:gene:2004}. For different sets of parameters, we checked
numerically that in these two cases our formulas are consistent with the
formulas based on modified Bessel functions derived in \citet{Zack:gene:2004}.
}

\section{Joint Distribution of $X(t)$ and $S(t)$}
\label{sec:jointXS}

Let us first work out the details the formula for $P_0(X(t)\in dx,
S(t)=0)$. Fix $0<s<t$.  Given $M(t)=s$, random variable $X(t)$
has a normal distribution with mean 0 and variance $\sigma^2s$, because Markov
Chain $S(\cdot)$ and Brownian Motion $B(\cdot)$ are independent processes.
Let $\phi(\cdot,\sigma^2)$ denote the pdf  of a normal random variable with
mean zero and variance $\sigma^2$. Then we get that
\begin{equation*}
P_0(X(t)\in dx,S(t)=0,M(t)\in \dif s)=\phi(x,\sigma^2s)p_{00}(s,t)dxds.
\end{equation*}
Now, recall also that given $S(0)=0$, random variable $M(t)$ has an atom (with
weight $e^{-\lambda_0t}$ at $s=t$). Therefore, when we integrate $s$ out of
the joint distribution of $X(t)$, $S(t)$ and $M(t)$, we get that
\begin{equation}\label{h00}
h_{00}(x,t)=P_0(X(t)\in dx,S(t)=0)/dx=e^{-\lambda_0t}\phi(x,\sigma^2t)+\int_0^t \phi(x,\sigma^2s)p_{00}(s,t)ds.
\end{equation}
In a similar fashion, one can show that for $i=1,2$
\begin{equation}\label{h0i}
h_{0i}(x,t)=P_0(X(t)\in dx,S(t)=i)/dx=\int_0^t \phi(x,\sigma^2s)p_{0i}(s,t)ds.
\end{equation}

When $S(0)=1$, the distribution of random variable $X(t)$ has an atom at $x=0$
(if $R_0>t$, that is, the Markov chain stays in state~1 till time $t$). Taking
this into an account we have the following formulas:
\begin{equation}\label{h1i}
h_{1i}(x,t)=P_1(X(t)\in dx,S(t)=i)/dx=\int_0^t \phi(x,\sigma^2s)p_{1i}(s,t)ds,\quad \mbox{ if } x\neq 0 \mbox{ and } i=0,1,2,
\end{equation}
and
\begin{equation*}
P_1(X(t)=0,S(t)=1)=e^{-\lambda_1t}.
\end{equation*}

Similarly,
\begin{equation}\label{h2i}
h_{2i}(x,t)=P_2(X(t)\in dx,S(t)=i)/dx=\int_0^t \phi(x,\sigma^2s)p_{2i}(s,t)ds,\quad \mbox{ if } x\neq 0 \mbox{ and } i=0,1,2,
\end{equation}
and
\begin{equation*}
P_2(X(t)=0,S(t)=2)=e^{-\lambda_2t}.
\end{equation*}

{
It is not essential to use one-dimensional Brownian Motion for these
derivations but it simplifies our presentation's notation.
If one does want to consider a Brownian Motion of $d$-dimension, then all we
need to do is to substitute the one-dimensional normal pdf in formulas
(\ref{h00})--(\ref{h2i}) by the $d$-dimensional normal density with mean
zero and covariance matrix $\sigma^2 I_d$, where $I_d$ is the
$d$-dimensional identity matrix. Of course, in this case $x$ is a vector in the $d$-dimensional space,
not a scalar. In fact, later when we run
simulations and analyze real-world data we will use the two-dimensional setup.
}

\section{Likelihood Estimation with Forward Algorithm}
\label{sec:like}

Assume that we observe the MRH process $X(t)$  at times $0=t_0<t_1<\dots<t_n$.
Let $\X = \big(X_1, X_2, \dots, X_n\big)$, where $X_k=X(t_i)-X(t_{i-1})$,
$i=1,\dots,n$ are the observed increments of the MHR process.
Let $\St=\big(S(0),S(t_1),\dots,S(t_n)\big)$ be the corresponding states of
the the continuous-time Markov Chain, and $\Delta_i=t_i-t_{i-1}$,
$i=1,\dots,n$.

The location-state process $\{X(t), S(t)\}$ is Markov, so the likelihood
function of $(\X,\St)$  is available in closed-form. More specifically,
it is given by
\begin{equation}\label{Full*L}
  L(\X,\St,{\boldsymbol\theta}) = \nu(S(0))\prod_{i=1}^n f\big( X_i, S(t_i)| S(t_{i-1}), \Delta_i, {\boldsymbol\theta}\big),
\end{equation}
where
\begin{equation}\label{eq:f}
  f\big( x, u | v, t, {\boldsymbol\theta}\big)
  =
  \begin{cases}
    0                        & v\neq u,\       x=0,\\
    0                        & v = u = 0,\     x=0,\\
    e^{-\lambda_1 t}         & v = u = 1,\     x=0,\\
    e^{-\lambda_2 t}         & v = u = 2,\     x=0,\\
    h_{ij}\big( x, t\big) & v = i,\ u = j,\ x\neq 0,
  \end{cases}
\end{equation}
$x\in\mathbf{R}$, $u,v=0,1,2$, $t>0$, and  ${\boldsymbol\theta}=(\lambda_0,\lambda_1,\lambda_2,p_1,\sigma)$.

The distribution of the increments of the MRH process is a mixture of
absolutely continuous and discrete distributions. Therefore, in order
to construct the likelihood function we have to use the Radon--Nikodym
derivative of the probability distribution relative to a dominating
measure that includes an atom at $x=0$. That explains the special sets
of formulas in the case when $x=0$.

Now, if the state vector $S_t$ is not observed, then obviously the
likelihood of the increment vector $\X$ can be computed using
\begin{equation*}
  L(\X,{\boldsymbol\theta}) = \sum_{s_0,\dots, s_n}L(\X,(s_0,\dots,s_n),{\boldsymbol\theta}),
\end{equation*}
where the summation is taken over all possible trajectories of $\St$.
However, this formula is not practical since the number of trajectories grows
exponentially as sample size $n\to \infty$. This difficulty is addressed with
help of the forward algorithm.

First, we need to introduce forward variables:
\begin{equation}\label{alpha*k}
\alpha(\X_k,s_k,{\boldsymbol\theta}) = \sum_{s_0,\dots, s_{k-1}}\nu(s_0)\prod_{i=1}^k f\big( X_i, s_i | s_{i-1}, \Delta_i, {\boldsymbol\theta}\big),
\end{equation}
where $\X_k =\big(X_1,X_2,\dots,X_k\big)$, and $1\leq k\leq n$.
Then one can show that
\begin{equation}\label{alpha*k*formula}
\alpha(\X_{k+1},s_{k+1},{\boldsymbol\theta})= \sum_{s_{k}}f\big( X_{k+1}, s_{k+1} | s_{k}, \Delta_{k+1}, {\boldsymbol\theta}\big)\alpha(\X_k,s_k,{\boldsymbol\theta}).
\end{equation}
{
That is, for every $k$  we have three forward variables. To get one
$k+1$st forward variable we need to calculate three transitional
values in~\eqref{eq:f}, multiply each $k$th forward variable by an
appropriate transitional value, and finally sum up these three
quantities. The bottom line is that the transition from
$\alpha(\X_{k},s_{k},{\boldsymbol\theta})$ to
$\alpha(\X_{k+1},s_{k+1},{\boldsymbol\theta})$ for each $k$ requires a
constant (independent of $k$) number of operations.
Since
 \begin{equation*}
  L(\X,{\boldsymbol\theta}) = \sum_{s_n}\alpha(\X_n,s_n,{\boldsymbol\theta}),
\end{equation*}
we get an algorithm that finds $L(\X,{\boldsymbol\theta})$ with computational complexity that is linear with respect to sample size $n$.
}

The next step is to modify the forward variables to address the
underflow problem. The problem is that for large $k$ forward variables
$\alpha(\X_{k},s_{k},{\boldsymbol\theta})$ might be numerically
indistinguishable from zero. To resolve this issue the following
normalized forward variables are employed:
\begin{equation}\label{normalized*alpha*k*formula}
\bar{\alpha}(\X_k,s_k,{\boldsymbol\theta})=\frac{\alpha(\X_k,s_k,{\boldsymbol\theta})}{L(\X_k,{\boldsymbol\theta})},
\end{equation}
where $L(\X_k,{\boldsymbol\theta})=\sum_{s_k} \alpha(\X_k,s_k,{\boldsymbol\theta})$, the likelihood of vector $\X_k$.
Then (\ref{alpha*k*formula}) immediately implies that the normalized forward variables satisfy the following equation:
\begin{equation*}\label{alpha*bar*k}
\bar{\alpha}(\X_{k+1},s_{k+1},{\boldsymbol\theta})=\frac{L(\X_{k},{\boldsymbol\theta})}{L(\X_{k+1},{\boldsymbol\theta})}
\sum_{s_{k}}f\big( X_{k+1}, s_{k+1} | s_{k},\Delta_{k+1},{\boldsymbol\theta}\big)\bar{\alpha}(\X_k,s_k,{\boldsymbol\theta}).
\end{equation*}
If for $0\leq k\leq n-1$  we define
\begin{equation*}
d(\X_{k+1},{\boldsymbol\theta})=\frac{L(\X_{k+1},{\boldsymbol\theta})}{L(\X_k,{\boldsymbol\theta})},
\end{equation*}
then one can easily verify that
\begin{equation*}\label{d*k}
d(\X_{k+1},{\boldsymbol\theta})=\sum_{s_{k+1}}\sum_{s_{k}}f\big( X_{k+1}, s_{k+1} | s_{k}, \Delta_{k+1},{\boldsymbol\theta}\big)\bar{\alpha}(\X_k,s_k,{\boldsymbol\theta}).
\end{equation*}

Here is the normalized version of the forward algorithm.
\begin{enumerate}
\item For observed $\X$ and given parameter vector ${\boldsymbol\theta}$, compute
  $f\big( X_{k+1}, s_{k+1} | s_{k},\Delta_{k+1},{\boldsymbol\theta}\big)$ for all possible
  pairs $(s_k, s_{k+1})$, $k=0,\dots,n-1$.
\item Base case:  $\bar{\alpha}(\X_0,s_0,{\boldsymbol\theta}) = \nu(s_0)$, where $s_0=0,1,2$.
\item Induction: for $s_{k+1}=0,1,2$ compute $\bar{\alpha}(\X_{k+1},s_{k+1},{\boldsymbol\theta})$ using
\begin{equation*}
\bar{\alpha}(\X_{k+1},s_{k+1},{\boldsymbol\theta})=\frac{1}{d(\X_{k+1},{\boldsymbol\theta})}
\sum_{s_{k}}f\big( X_{k+1}, s_{k+1} | s_{k},\Delta_{k+1},{\boldsymbol\theta}\big)\bar{\alpha}(\X_k,s_k,{\boldsymbol\theta}).
\end{equation*}
and
\begin{equation*}
d(\X_{k+1},{\boldsymbol\theta})=\sum_{s_{k+1}}\sum_{s_{k}}f\big( X_{k+1}, s_{k+1} | s_{k}, \Delta_{k+1},{\boldsymbol\theta}\big)\bar{\alpha}(\X_k,s_k,{\boldsymbol\theta}).
\end{equation*}
\item Termination: $\log L(\X,{\boldsymbol\theta}) = \sum_{k=1}^n \log d(\X_{k},{\boldsymbol\theta})$.
\end{enumerate}

{
This algorithm can be easily adapted to a situation when some states
are completely observed or partially observed. For example, accelerometer
data might be used to infer when an animal is moving or not, and
direct inspection of a kill-site can confirm handling. If state $s_{k}$ is known,
then first calculate three $k$th forward
variables as usual. Next, set the two forward variables with
unobservable states to zero. After that just continue the forward
algorithm in the normal fashion until the next location where additional
information on the state is available. If at $k$th location only one state
is excluded, then we have to set only one forward variable to zero.
}

\section{Simulation and Data Analysis}\label{sec:sim}

\begin{figure}[tbp]
\centering
\includegraphics[angle=-90, scale=.5]{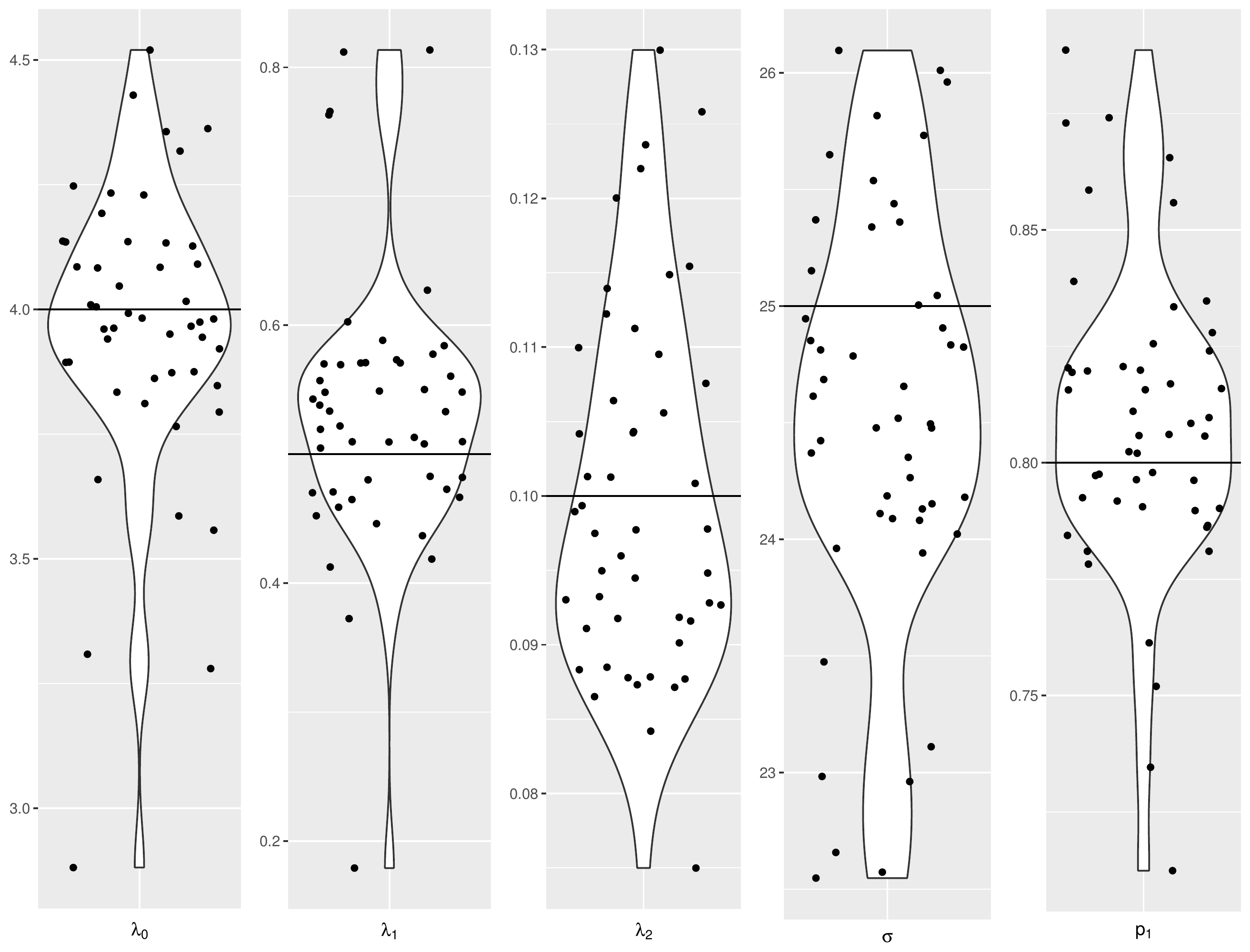}
\caption{Violin plots of the maximum likelihood estimates from 49 replicates
  using the forward algorithm. The horizontal bar in each panel is the true
  parameter value.}
\label{fig:mle}
\end{figure}

{
We ran a small simulation to demonstrate that the forward algorithm successfully recovers the model
parameters. The true parameter values were set to be
$\lambda_0 = 4$, $\lambda_1 = 0.5$, $\lambda_2 = 0.1$, $p_1 = 0.8$, and
$\sigma = 25$. The simulation was small because the
computation of the maximum likelihood estimator is very demanding.
Evaluation of the terms in Theorems~\ref{thm:S0=0}--\ref{thm:S0=1}
involves infinite series that are computationally intensive; evaluation of the
terms in the likelihood in Section~\ref{sec:jointXS} is very expensive because
functions in~\eqref{eq:f} are numerical integrals of $p_{ij}(s,t)$.
We generated $49$ two-dimensional datasets on a time grid from~0
to~4000, with increment 20, so the resulting series $\X$ is of length~200.

Figure~\ref{fig:mle} presents the violin plots of the likelihood estimates of
the 49~replicates in comparison to the true values of the five parameters.
Violin plots are similar to box plots with a rotated kernel density plot on
each side, which show more information about the data than box plots.
The horizontal bars in the panels are the true parameter values.
For each parameter, the true value lies in the bulk part of the violin plot,
indicating that the true parameters are recovered well by the likelihood
estimates in this small scale simulation study.

We next applied the proposed model to the data from the same mountain lion
analyzed by \citet{Yan:etal:2014} and \citet{Pozd:etal:2017}. This
mount lion was a mature female in the Gros Ventre Mountain Range near
Jackson Wyoming tracked with a GPS collar from 2009 to 2012.
The collar was designed to collect a fix every 8 hours but the actual
sampling times were irregular with sampling intervals having standard
deviation 6.45~hours, ranging from 0.5 hours to 120~hours.
Mountain lions behave differently in the summer and in the winter, so we
focused on the summer of 2012, a total of 389 observations spanning
from June~1 to August~31, which makes our results not directly
comparable to existing analyses \citet{Yan:etal:2014, Pozd:etal:2017}.
Field personnel determined that some of the sites were places where the
mountain lion consumed a prey item. She typically remained within
250~m of a kill site while it was considered to be ``handling'',
which is different from shorter, resting periods.
To allow for GPS measurement error, we rounded the locations to the
nearest 100~meters.

The maximum likelihood estimates of the MRH model parameters are:
$\hat\lambda_0 = 9.25$/hour, $\hat\lambda_1 = 2.49$/hour,
$\hat\lambda_2 = 0.19$/hour, $\hat\sigma = 1.28$km/hour$^{1/2}$,
and $\hat p = 0.70$. That is, on average, the mountain lions stays for
0.11, 0.40, and 5.1 hours in the moving, resting, and handling states.
When moving, the mobility parameter is 1.28km/hour$^{1/2}$. This means that 
if the mountain lion moves without stopping for one hour, the average deviation 
from the initial position in terms of northing and easting values is 1.28 km.   
When she stopped moving, she went into resting with probability 0.70 and
handling with probability 0.30, respectively.
For comparison, we also fitted the moving-resting process to the same
data, and the maximum likelihood estimates of the parameters are
$\hat\lambda_0 = 0.66$/hour, $\hat\lambda_1 = 0.27$/hour, and
$\hat\sigma = 0.59$km/hour$^{1/2}$. Because there is no handling state, the
average durations in both moving and resting are estimated longer.
Consequently, the mobility parameter estimate is much lower,
almost halved, because the animal was assumed to be moving longer.
We also fitted the BBMM of \citep{Horn:etal:2007} with the original,
non-rounded data and the GPS measurement error standard deviation
fixed at 0.02km. The BM mobility parameter estimate is even lower,
0.42km/hour$^{1/2}$, as the animal was assumed to be always moving.
}

\section{Concluding Remarks}\label{sec:conc}

{ 
The results on occupation times obtained in the paper have their own value and
can be used for other applications, such as quality control. Indeed, the
continuous-time Markov Chain $S(t)$ can be viewed as a telegraph process with
two off states. These two states will correspond two different types of
breakdown that require different time for repair.
The results in Theorems~\ref{thm:S0=0} and~\ref{thm:S0=1} can be
easily generalized to cover $k$~motionless states instead of just two.
The only difference is that, instead of binomial distribution and
convolutions of two gamma distributions, we will have multinomial
distribution and convolutions of $k$ gammas.

The methodology developed in Sections~\ref{M(t)*when*S(0)=0*section}
and \ref{M(t)*when*S(0)=1*section} works even if the holding times are not
exponentially distributed, which is an advantage of our approach.
If we want to keep the Markov property, then all holding times {\it must}
have exponential distributions. The memoryless distribution might be not
appropriate for some species that follow a cyclic daily routine.
Nonetheless, if animals under observation do not exhibit a daily periodic behavior
(like mountain lions), then using an exponential distribution is acceptable.
The behavior of these animals is subject to interruptions
that can cut their time spent in a particular activity. For example,
handling might be interrupted by a more dominate predator who drives the lion
off her kill before she is finished with it.

A different (from exponential) distribution should be used for species with a
periodic routine. One interesting possibility is to employ stable
distributions (for example, L\'{e}vy distribution).
Because a linear combination of two independent random variables with a stable
distribution has the same distribution, up to location and scale parameters,
the formulas in Theorems~\ref{thm:S0=0} and~\ref{thm:S0=1} will be even nicer.
The drawback is that the state process is then semi-Markov, and, as a
result, the likelihood inferences from standard HMM tools are not available.
Nevertheless, this still might be of interest for practitioners in
ecological science, because estimation can be done via alternative methods
such as the composite likelihood estimation \citep{Lind:comp:1988}.
}

\bibliographystyle{mcap}
\bibliography{MRH}

\begin{thebibliography}{23}
\newcommand{\enquote}[1]{``#1''}
\expandafter\ifx\csname natexlab\endcsname\relax\def\natexlab#1{#1}\fi

\bibitem[{Bshouty et~al.(2012)Bshouty, Di~Crescenzo, Martinucci and
  Zacks}]{Bshouty:etal:2012}
D.~Bshouty, A.~Di~Crescenzo, B.~Martinucci and S.~Zacks (2012).
\newblock \enquote{Generalized telegraph process with random delays.}
\newblock {\em Journal of Applied Probability\/} {\bf 49}, 850--865.

\bibitem[{Capp{\'e} et~al.(2005)Capp{\'e}, Moulines and
  Ryd{\'e}n}]{Capp:etal:Infe:2005}
O.~Capp{\'e}, E.~Moulines and T.~Ryd{\'e}n (2005).
\newblock {\em Inference in Hidden {M}arkov Models\/}.
\newblock Springer.

\bibitem[{Codling et~al.(2008)Codling, Plank and Benhamou}]{Codling:etal:2008}
E.~A. Codling, M.~J. Plank and S.~Benhamou (2008).
\newblock \enquote{Random walk models in biology.}
\newblock {\em Journal of The Royal Society Interface\/} {\bf 5}, 813--834.

\bibitem[{Di~Crescenzo(2001)}]{DiCrescenzo:2001}
A.~Di~Crescenzo (2001).
\newblock \enquote{On random motions with velocities alternating at
  {E}rlang-distributed random times.}
\newblock {\em Advances in Applied Probability\/} {\bf 33}, 690--701.

\bibitem[{Di~Crescenzo et~al.(2014)Di~Crescenzo, Martinucci and
  Zacks}]{DiCrescenzo:etal:2014}
A.~Di~Crescenzo, B.~Martinucci and S.~Zacks (2014).
\newblock \enquote{On the geometric brownian motion with alternating trend.}
\newblock In C.~Perna and M.~Sibillo (eds.), {\em Mathematical and Statistical
  Methods for Actuarial Sciences and Finance\/}, pp. 81--85. Dordrecht:
  Springer.

\bibitem[{Di~Crescenzo and Pellerey(2002)}]{DiCrescenzo:Pellerey:2002}
A.~Di~Crescenzo and F.~Pellerey (2002).
\newblock \enquote{On prices' evolutions based on geometric telegrapher's
  process.}
\newblock {\em Applied Stochastic Models in Business and Industry\/} {\bf 18},
  171--184.

\bibitem[{Di~Crescenzo and Zacks(2015)}]{DiCrescenzo:Zhacs:2015}
A.~Di~Crescenzo and S.~Zacks (2015).
\newblock \enquote{Probability law and flow function of {B}rownian motion
  driven by a generalized telegraph process.}
\newblock {\em Methodology and Computing in Applied Probability\/} {\bf 17},
  761--780.

\bibitem[{Horne et~al.(2007)Horne, Garton, Krone and Lewis}]{Horn:etal:2007}
J.~S. Horne, E.~O. Garton, S.~M. Krone and S.~Lewis, J (2007).
\newblock \enquote{Analyzing animal movements using {B}rownian bridges.}
\newblock {\em Ecology\/} {\bf 88}, 2354--2363.

\bibitem[{Kolesnik and Ratanov(2013)}]{Kolesnik:Ratanov:2013}
A.~D. Kolesnik and N.~Ratanov (2013).
\newblock {\em Telegraph processes and option pricing\/}.
\newblock Springer Briefs in Statistics. Springer, Heidelberg.

\bibitem[{Lindsay(1988)}]{Lind:comp:1988}
B.~G. Lindsay (1988).
\newblock \enquote{Composite likelihood methods.}
\newblock {\em Contemporary Mathematics\/} {\bf 80}, 221--239.

\bibitem[{Mathai(1982)}]{Math:1982}
A.~Mathai (1982).
\newblock \enquote{The storage capacity of a dam with gamma type inputs.}
\newblock {\em Annals of Institute of Statistical Mathematics\/} {\bf 34},
  591--597.

\bibitem[{Moschopoulos(1985)}]{Mosc:1985}
P.~Moschopoulos (1985).
\newblock \enquote{The distribution of the sum of independent gamma random
  variables.}
\newblock {\em Annals of Institute of Statistical Mathematics\/} {\bf 37},
  541--544.

\bibitem[{Othmer et~al.(1988)Othmer, Dunbar and Alt}]{Othmer:1988}
H.~G. Othmer, S.~R. Dunbar and W.~Alt (1988).
\newblock \enquote{Models of dispersal in biological systems.}
\newblock {\em Journal of Mathematical Biology\/} {\bf 26}, 263--298.

\bibitem[{Perry et~al.(1999)Perry, Stadje and Zacks}]{Perr:Stad:Zack:firs:1999}
D.~Perry, W.~Stadje and S.~Zacks (1999).
\newblock \enquote{First-exit times for increasing compound processes.}
\newblock {\em Communications in Statistics: Stochastic Models\/} {\bf 15},
  977--992.

\bibitem[{Pozdnyakov et~al.(2017)Pozdnyakov, Elbroch, Labarga, Meyer and
  Yan}]{Pozd:etal:2017}
V.~Pozdnyakov, L.~Elbroch, A.~Labarga, T.~Meyer and J.~Yan (2017).
\newblock \enquote{Discretely observed {B}rownian motion governed by telegraph
  process: estimation.}
\newblock {\em Methodology and Computing in Applied Probability\/} {t}o appear.

\bibitem[{Preisler et~al.(2004)Preisler, Ager, Johnson and
  Kie}]{Prei:etal:mode:2004}
H.~K. Preisler, A.~A. Ager, B.~K. Johnson and J.~G. Kie (2004).
\newblock \enquote{Modeling animal movements using stochastic differential
  equations.}
\newblock {\em Environmetrics\/} {\bf 15}, 643--657.

\bibitem[{Sericola(2000)}]{Seri:2000}
B.~Sericola (2000).
\newblock \enquote{Occupation times in markov processes.}
\newblock {\em Communications in Statistics. Stochastic Models\/} {\bf 16},
  479--510.

\bibitem[{Stadje and Zacks(2004)}]{Stad:Zack:tele:2004}
W.~Stadje and S.~Zacks (2004).
\newblock \enquote{Telegraph processes with random velocities.}
\newblock {\em Journal of Applied Probability\/} {\bf 41}, 665--678.

\bibitem[{Tilles and Petrovskii(2016)}]{Tilles:Petr:2016}
P.~F.~C. Tilles and S.~V. Petrovskii (2016).
\newblock \enquote{How animals move along? exactly solvable model of
  superdiffusive spread resulting from animal's decision making.}
\newblock {\em Journal of mathematical biology\/} {\bf 73}, 227--55.

\bibitem[{Yan et~al.(2014)Yan, Chen, Lawrence-Apfel, Ortega, Pozdnyakov,
  Williams and Meyer}]{Yan:etal:2014}
J.~Yan, Y.-W. Chen, K.~Lawrence-Apfel, I.~Ortega, V.~Pozdnyakov, S.~Williams
  and T.~Meyer (2014).
\newblock \enquote{A moving-resting process with an embedded {B}rownian motion
  for animal movements.}
\newblock {\em Population Ecology\/} {\bf 56}, 401--415.

\bibitem[{Yan and Pozdnyakov(2016)}]{Rpkg:smam}
J.~Yan and V.~Pozdnyakov (2016).
\newblock {\em {smam}: Statistical Modeling of Animal Movements\/}.
\newblock R package version 0.3-0.

\bibitem[{Zacks(2004)}]{Zack:gene:2004}
S.~Zacks (2004).
\newblock \enquote{Generalized integrated telegraph processes and the
  distribution of related stopping times.}
\newblock {\em Journal of Applied Probability\/} {\bf 41}, 497--507.

\bibitem[{Zacks(2012)}]{Zacks:2012}
S.~Zacks (2012).
\newblock \enquote{Distribution of the total time in a mode of an alternating
  renewal process with applications.}
\newblock {\em Sequential Analysis\/} {\bf 31}, 397--408.

\end{thebibliography}

\end{document}